\documentclass[conference,letterpaper]{IEEEtran}
\IEEEoverridecommandlockouts
\usepackage{cite}
\usepackage{amsmath,amssymb,amsfonts}
\usepackage{graphicx}
\usepackage{textcomp}
\usepackage{xcolor}
\usepackage{balance}
\usepackage{epsfig}
\usepackage{epstopdf}

\usepackage[linesnumbered,ruled,vlined]{algorithm2e}
\usepackage{float}
\usepackage{subfig}
\usepackage{times}
\usepackage{url}
\usepackage{microtype}

\usepackage{bm}
\usepackage{dsfont} 

\usepackage{tikz}
\usetikzlibrary{arrows.meta,positioning,fit,calc}

\def\BibTeX{{\rm B\kern-.05em{\sc i\kern-.025em b}\kern-.08em
    T\kern-.1667em\lower.7ex\hbox{E}\kern-.125emX}}

\begin{document}

\title{Reduced-Feedback Hybrid Precoding for Wideband mmWave MIMO-OFDM Systems via Hierarchical Interpolation \vspace{-0.1in} 
\thanks{This work was supported in part by the National Science and Technology Council (NSTC) of Taiwan under Grant 113-2926-I-001-502-G, 113-2221-E-110-059-MY3, and NSTC 115-2640-E-110-003, and also in part by the Sixth Generation Communication and Sensing Research Center funded by the Higher Education SPROUT Project, the Ministry of Education of Taiwan.}
}
\author{\IEEEauthorblockN{Po-Heng Chou$^{1,3}$, Jia-Qing Lin$^{2}$, Wei-Chen Lin$^{2}$, Wan-Jen Huang$^{2}$, and Ronald Y. Chang$^{1,4}$}
\IEEEauthorblockA{
$^{1}$Research Center for Information Technology Innovation (CITI), Academia Sinica (AS), Taipei 11529, Taiwan\\
$^{2}$Institute of Communication Engineering (ICE), National Sun Yat-sen University (NSYSU), Kaohsiung 80424, Taiwan\\
$^{3}$Bradley Department of Electrical and Computer Engineering (ECE), Virginia Tech (VT), Alexandria, VA 22305, USA\\
$^{4}$Department of Electrical Engineering, National Cheng Kung University (NCKU), Tainan 701401, Taiwan\\
E-mails: d00942015@ntu.edu.tw, elinqing514@gmail.com, wilson20020501@gmail.com,\\ wjhuang@faculty.nsysu.edu.tw, rchang@citi.sinica.edu.tw\vspace{-0.2in}
}
}

\maketitle
\begin{abstract}
In this paper, we propose a feedback-efficient hybrid precoding framework for wideband millimeter-wave (mmWave) multiple-input multiple-output orthogonal frequency-division multiplexing (MIMO-OFDM) systems. To reduce the hardware cost of radio-frequency (RF) chains and the feedback burden associated with frequency-selective digital precoding, frequency-flat analog precoders and combiners are first constructed from the dominant angle-of-arrival (AoA) and angle-of-departure (AoD) directions associated with the strongest propagation paths. For digital precoding, a shared quantized codebook is designed using the Lloyd algorithm, and a binary-search-based hierarchical interpolation method is developed to identify codeword switching locations between adjacent pilot subcarriers. By avoiding exhaustive codeword comparisons over all intermediate subcarriers, the proposed method reduces online search complexity and enables compact digital-precoder feedback. For the considered system configuration, the proposed method achieves a worst-case feedback reduction of approximately $98.34\%$ relative to full-grid per-subcarrier feedback. Simulation results further demonstrate competitive spectral efficiency and bit error rate (BER) performance, low computational cost, and robustness under imperfect effective channel state information (CSI).
\end{abstract}


\section{Introduction}

Millimeter-wave (mmWave) multiple-input multiple-output orthogonal frequency-division multiplexing (MIMO-OFDM) is a promising physical-layer architecture for future wideband wireless systems because large antenna arrays can provide substantial beamforming and spatial multiplexing gains~\cite{Heath2016}. However, fully digital precoding requires a dedicated radio-frequency (RF) chain for each antenna, resulting in high hardware cost and power consumption. Hybrid analog--digital precoding alleviates this limitation by combining a high-dimensional analog precoder with lower-dimensional digital precoders~\cite{Ayach2014,Gao2016}.

In wideband MIMO-OFDM systems, the analog precoder is generally shared across subcarriers, whereas the digital precoder must adapt to frequency-selective channel variations. Reporting an independently selected digital precoder for every subcarrier therefore causes the feedback overhead and online codeword-search complexity to increase linearly with the number of subcarriers~\cite{Alkhateeb2016,Lin2018TSP}. Limited-feedback methods exploit the correlation among adjacent subcarriers to reduce the number of explicitly reported precoders.

Early feedback-reduction approaches mainly relied on clustering and interpolation. Clustering-based methods assign a common precoder to a group of adjacent subcarriers~\cite{Wu2009}, whereas interpolation-based methods determine the precoders of unreported subcarriers from a sparse set of anchor subcarriers~\cite{Choi2005,Pande2007,He2011}. Although these methods reduce feedback overhead, their subcarrier partitions or interpolation structures are generally fixed in advance and may not efficiently track changes in the preferred precoding codeword within an interval.

Codebook-based hybrid precoding provides another practical approach by representing analog or digital precoders using indices from a codebook shared between the transmitter and receiver. Existing designs include frequency-selective hybrid codebooks~\cite{Alkhateeb2016}, non-uniform quantization codebooks~\cite{Chen2019TCom}, group-and-codebook-based hybrid precoding~\cite{Lin2023TWC}, and partial discrete Fourier transform (DFT) codebooks for mmWave massive MIMO systems~\cite{Shi2024}. These methods improve quantization or reduce precoder-reporting overhead, but they do not directly address adaptive codeword assignment across a large number of frequency-selective subcarriers.

More recent studies have applied machine learning to CSI compression and hybrid precoding. End-to-end learning frameworks jointly optimize pilot design, CSI feedback, and hybrid beamforming~\cite{Yang2024}, while implicit CSI feedback and beamforming~\cite{Xue2023WCL} reduce explicit CSI acquisition requirements. TypeII-CsiNet further integrates an autoencoder-based feedback architecture with a standardized Type-II codebook~\cite{Sang2024TypeIICsiNet}. However, these learning-based methods generally require representative training data, neural-network model storage, and model retraining or adaptation when the channel distribution or hardware configuration changes. In contrast, the proposed method does not require a neural inference model during online precoder selection; its Lloyd codebook is generated offline using conventional vector quantization.

Motivated by these observations, we develop a reduced-feedback hybrid precoding framework using analog beamforming based on the dominant angle-of-arrival (AoA) and angle-of-departure (AoD) directions, a shared Lloyd-trained digital codebook, and binary-search-based hierarchical interpolation. The proposed search identifies a switching location between neighboring pilot codewords under a single-transition model, reducing the number of gain evaluations from linear to logarithmic scaling with the pilot spacing.

The main contributions are summarized as follows:
\begin{itemize}
\setlength{\itemsep}{0pt}
\setlength{\parskip}{0pt}
\setlength{\parsep}{0pt}

    \item We develop a reduced-feedback hybrid precoding framework
    combining dominant-path analog beamforming with codebook-based
    digital precoding.

    \item We propose a binary-search-based hierarchical interpolation
    method to identify codeword switching locations without exhaustive
    subcarrier-wise comparisons.

    \item We characterize the online search complexity and feedback
    overhead of the proposed method.

    \item Simulations demonstrate competitive spectral efficiency and bit error rate (BER), low runtime, and robustness under imperfect CSI.
\end{itemize}

\section{System and Channel Models}
\label{sec:system_model}

\begin{figure}[t]
\centering
\includegraphics[width=0.5\textwidth]{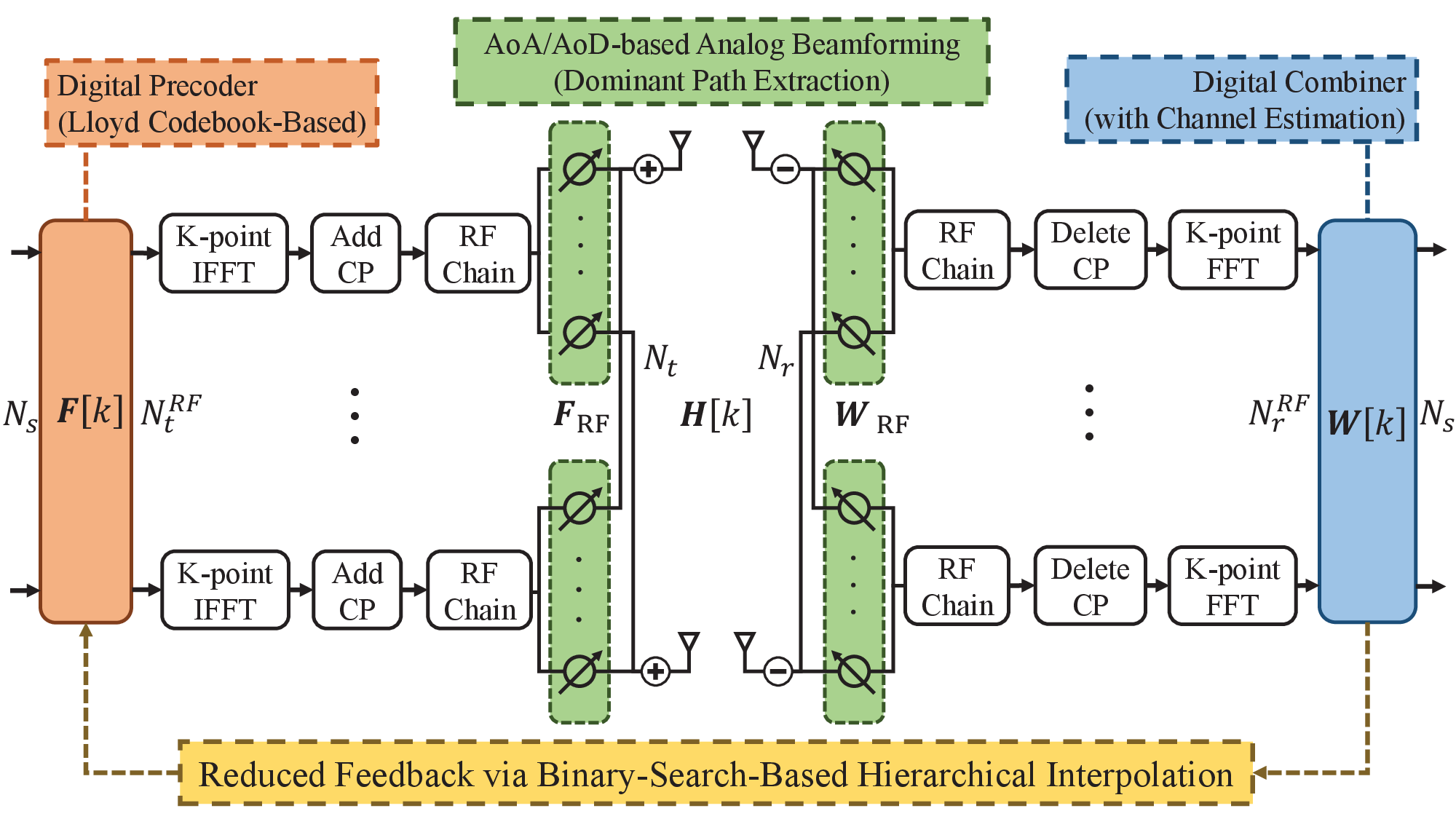}
\caption{\small Reduced-feedback hybrid precoding framework with dominant-path analog beamforming, Lloyd-codebook-based digital precoding, and binary-search-based hierarchical interpolation.}
\label{System_Model}
\end{figure}

Consider a single-user wideband mmWave MIMO-OFDM downlink system, as illustrated in Fig.~\ref{System_Model}. The transmitter is equipped with $N_t$ antennas and $N_t^{\mathrm{RF}}$ RF chains, while the receiver employs $N_r$ antennas and $N_r^{\mathrm{RF}}$ RF chains, where
$N_s \leq N_t^{\mathrm{RF}} \leq N_t$ and
$N_s \leq N_r^{\mathrm{RF}} \leq N_r$.
The transmitter sends $N_s$ spatial streams over $K$ OFDM subcarriers. The analog precoder and combiner are frequency-flat and shared across all subcarriers, whereas the digital precoder and combiner are frequency-selective. Throughout this paper, $(\cdot)^H$ denotes the Hermitian transpose, $\|\cdot\|_2$ and $\|\cdot\|_F$ denote the Euclidean and Frobenius norms, respectively, and $\mathbb{E}[\cdot]$ denotes expectation.

Let $\bm{s}[k]\in\mathbb{C}^{N_s\times1}$ denote the transmitted
symbol vector on the $k$-th subcarrier, with
$\mathbb{E}[\bm{s}[k]\bm{s}^{H}[k]]
=(P/N_s)\bm{I}_{N_s}$, where $P$ denotes the transmit power
allocated to each subcarrier and $\bm{I}_{N}$ denotes the
$N\times N$ identity matrix.
The symbol vector is first processed by the digital baseband precoder
$\bm{F}[k] \in \mathbb{C}^{N_t^{\mathrm{RF}} \times N_s}$.
The resulting frequency-domain signals are converted into time-domain OFDM signals through a $K$-point inverse fast Fourier transform (IFFT), followed by cyclic-prefix insertion and RF upconversion.

A frequency-flat analog precoder
$\bm{F}_{\mathrm{RF}} \in
\mathbb{C}^{N_t \times N_t^{\mathrm{RF}}}$
is subsequently applied through a phase-shifter network.
The baseband-equivalent transmitted signal on the $k$-th subcarrier is $\bm{x}[k]
=
\bm{F}_{\mathrm{RF}}\bm{F}[k]\bm{s}[k]$.
Each entry of the analog precoder satisfies
$|[\bm{F}_{\mathrm{RF}}]_{m,n}|^2=1/N_t$ and can be written as
$[\bm{F}_{\mathrm{RF}}]_{m,n}
=N_t^{-1/2}e^{j\theta^{(t)}_{m,n}}$,
where $\theta^{(t)}_{m,n}$ denotes the transmit-side phase shift applied by the analog phase-shifter network.

At the receiver, assuming perfect timing and frequency
synchronization, the received signal on the $k$-th subcarrier is
$\bm{z}[k]=\bm{H}[k]\bm{x}[k]+\bm{n}[k]$, where
$\bm{H}[k]\in\mathbb{C}^{N_r\times N_t}$ denotes the
frequency-domain channel matrix and
$\bm{n}[k]\sim\mathcal{CN}(\bm{0},\sigma_n^2\bm{I}_{N_r})$
denotes circularly symmetric complex Gaussian noise.

The receiver first applies a frequency-flat analog combiner
$\bm{W}_{\mathrm{RF}}
\in
\mathbb{C}^{N_r\times N_r^{\mathrm{RF}}}$.
Its entries satisfy
$|[\bm{W}_{\mathrm{RF}}]_{m,n}|^2=1/N_r$ and can be written as
$[\bm{W}_{\mathrm{RF}}]_{m,n}
=N_r^{-1/2}e^{j\theta^{(r)}_{m,n}}$,
where $\theta^{(r)}_{m,n}$ denotes the receive-side phase shift applied by the analog phase-shifter network.

After analog combining, RF downconversion, cyclic-prefix removal, and a $K$-point fast Fourier transform (FFT), a digital combiner
$\bm{W}[k]\in
\mathbb{C}^{N_r^{\mathrm{RF}}\times N_s}$
is applied.
The resulting output signal is
\begin{align}
\bm{y}[k]
=
\bm{W}^{H}[k]\bm{W}_{\mathrm{RF}}^{H}
\bigl(
\bm{H}[k]\bm{F}_{\mathrm{RF}}\bm{F}[k]\bm{s}[k]
+\bm{n}[k]
\bigr).
\label{received_signal}
\end{align}

\subsection{Wideband mmWave Channel Model}

We adopt a geometric frequency-selective mmWave channel with $L$ propagation paths.
The channel impulse response at delay tap $d$ is modeled as
\begin{align}
\bm{H}(d)
=
\sqrt{\frac{N_t N_r}{L}}
\sum_{\ell=1}^{L}
\alpha_{\ell}
p_{\mathrm{rc}}(dT_s-\tau_{\ell})
\bm{a}_r(\phi_{\ell})
\bm{a}_t^{H}(\theta_{\ell}),
\label{time_domain_channel}
\end{align}
where $\alpha_{\ell}$ is the complex gain of the $\ell$-th path,
$\tau_{\ell}$ is its propagation delay, $T_s$ is the sampling
interval, and $p_{\mathrm{rc}}(\cdot)$ denotes the pulse-shaping
response. The variables $\theta_{\ell}$ and $\phi_{\ell}$ denote the
AoD and AoA, respectively, while $\bm{a}_t(\theta_{\ell})$ and
$\bm{a}_r(\phi_{\ell})$ denote the corresponding transmit and receive
array-response vectors.

For analog beamforming, the path parameters $\{\alpha_{\ell},\tau_{\ell},\theta_{\ell},\phi_{\ell}\}_{\ell=1}^{L}$ are assumed available from preceding channel estimation and beam training. Effective-channel estimation errors are evaluated in Section~\ref{sec:simulation}.

The frequency-domain channel on the $k$-th subcarrier is obtained as
$\bm{H}[k]
=
\sum_{d=0}^{D-1}
\bm{H}(d)
e^{-j\frac{2\pi (k-1)d}{K}},
\quad
k=1,\ldots,K$,
where $D$ denotes the number of resolvable channel taps.

Both transceiver arrays are modeled as uniform linear arrays (ULAs) with inter-element spacing $d_{\mathrm{ant}}$. For an $N$-element ULA, the normalized array
response is
\begin{align}
\bm{a}_{N}(\vartheta)
=
\frac{1}{\sqrt{N}}
\left[
1,
e^{j\psi(\vartheta)},
\ldots,
e^{j(N-1)\psi(\vartheta)}
\right]^T,
\label{ula_response}
\end{align}
where
$\psi(\vartheta)=2\pi d_{\mathrm{ant}}\sin(\vartheta)/\lambda$,
with $\lambda$ denoting the carrier wavelength.
Thus, $\bm{a}_t(\theta)=\bm{a}_{N_t}(\theta)$ and
$\bm{a}_r(\phi)=\bm{a}_{N_r}(\phi)$.

Because the analog precoder and combiner are shared across all subcarriers, their application reduces the original channel to the lower-dimensional effective channel
\begin{align}
\bm{H}_{\mathrm{eff}}[k]
\triangleq
\bm{W}_{\mathrm{RF}}^{H}
\bm{H}[k]
\bm{F}_{\mathrm{RF}},
\label{effective_channel}
\end{align}
where
$\bm{H}_{\mathrm{eff}}[k]
\in
\mathbb{C}^{N_r^{\mathrm{RF}}\times N_t^{\mathrm{RF}}}$.
This effective channel is used for digital codeword selection and hierarchical interpolation in the following sections.

\section{Problem Formulation}

Using the effective channel defined in~\eqref{effective_channel}, the
received signal after analog and digital combining can be rewritten as
$\bm{y}[k]
=
\bm{H}_{\mathrm{eq}}[k]\bm{s}[k]
+
\widetilde{\bm{n}}[k]$,
where
$\bm{H}_{\mathrm{eq}}[k]
=
\bm{W}^{H}[k]\bm{H}_{\mathrm{eff}}[k]\bm{F}[k]
\in\mathbb{C}^{N_s\times N_s}$
denotes the equivalent channel and
$\widetilde{\bm{n}}[k]
=\bm{W}^{H}[k]\bm{W}_{\mathrm{RF}}^{H}\bm{n}[k]$
denotes the post-combining noise.
The post-combining noise covariance
$\bm{R}_{n}[k]\triangleq
\mathbb{E}[\widetilde{\bm{n}}[k]\widetilde{\bm{n}}^{H}[k]]$
is
\begin{align}
\bm{R}_{n}[k]
=
\sigma_n^2
\bm{W}^{H}[k]\bm{W}_{\mathrm{RF}}^{H}
\bm{W}_{\mathrm{RF}}\bm{W}[k].
\label{post_combining_noise_covariance}
\end{align}
Under Gaussian signaling, the average achievable spectral
efficiency is expressed as
\begin{align}
R
=
\frac{1}{K}
\sum_{k=1}^{K}
\log_2
\det
\left(
\bm{I}_{N_s}
+
\frac{P}{N_s}
\bm{R}_{n}^{-1}[k]
\bm{H}_{\mathrm{eq}}[k]
\bm{H}_{\mathrm{eq}}^{H}[k]
\right).
\label{spectral_efficiency}
\end{align}

The corresponding hybrid precoder--combiner design problem can be formulated as
\begin{subequations}
\label{optimization_problem}
\begin{align}
\max_{\substack{
\bm{F}_{\mathrm{RF}},\,\bm{W}_{\mathrm{RF}},\\
\{\bm{F}[k],\bm{W}[k]\}_{k=1}^{K}
}}
\quad
& R,
\label{optimization_objective}
\\
\mathrm{s.t.}\quad
&
\frac{1}{K}
\sum_{k=1}^{K}
\left\|
\bm{F}_{\mathrm{RF}}
\bm{F}[k]
\right\|_F^2
\leq
N_s,
\label{optimization_power_constraint}
\\
&
\left|
[\bm{F}_{\mathrm{RF}}]_{m,n}
\right|^2
=
\frac{1}{N_t},
\quad
\forall m,n,
\label{optimization_tx_constraint}
\\
&
\left|
[\bm{W}_{\mathrm{RF}}]_{m,n}
\right|^2
=
\frac{1}{N_r},
\quad
\forall m,n.
\label{optimization_rx_constraint}
\end{align}
\end{subequations}
Constraint~\eqref{optimization_power_constraint} specifies the average transmit-power budget of the hybrid precoder across all subcarriers, while \eqref{optimization_tx_constraint} and \eqref{optimization_rx_constraint} impose the constant-modulus constraints associated with the analog phase-shifter networks.

Problem~\eqref{optimization_problem} is non-convex due to the constant-modulus constraints and coupling between analog and digital beamformers. We therefore adopt a sequential low-complexity design: dominant-path analog beamforming, Lloyd-codebook-based pilot precoding, and hierarchical interpolation for the remaining subcarriers.

\section{Proposed Precoder and Codebook Design}
\label{sec:precoder_codebook}

Under the geometric mmWave channel model in~\eqref{time_domain_channel},
the frequency-domain channel matrices share the same physical AoA and
AoD directions across subcarriers, whereas their complex gains vary
with frequency. We exploit this common angular structure to construct
frequency-flat analog beamformers. The resulting lower-dimensional
effective channels are then used for codebook-based digital precoding.

\subsection{Analog Precoder and Combiner Design}

We first rank the propagation paths according to their aggregate energy
over the modeled delay taps, where
$\gamma_{\ell}
=
|\alpha_{\ell}|^2
\sum_{d=0}^{D-1}
|p_{\mathrm{rc}}(dT_s-\tau_{\ell})|^2$,
$\ell=1,\ldots,L$.

Let
$\mathcal{X}
=
\{\mathcal{X}^{(1)},\ldots,\mathcal{X}^{(L)}\}$
denote the path indices arranged in descending order of
$\gamma_{\ell}$, such that
$\gamma_{\mathcal{X}^{(1)}}\geq\cdots\geq
\gamma_{\mathcal{X}^{(L)}}$.
We assume
$L\geq\max\{N_t^{\mathrm{RF}},N_r^{\mathrm{RF}}\}$,
which is satisfied by the simulation configuration.

The transmit- and receive-side analog beams associated with the
$i$-th strongest path are selected as
$\bm{f}_i=\bm{a}_t(\theta_{\mathcal{X}^{(i)}})$,
$i=1,\ldots,N_t^{\mathrm{RF}}$, and
$\bm{w}_i=\bm{a}_r(\phi_{\mathcal{X}^{(i)}})$,
$i=1,\ldots,N_r^{\mathrm{RF}}$, respectively.
The frequency-flat analog precoder and combiner are then constructed as
$\bm{F}_{\mathrm{RF}}
=
[\bm{f}_1,\bm{f}_2,\ldots,\bm{f}_{N_t^{\mathrm{RF}}}]$
and
$\bm{W}_{\mathrm{RF}}
=
[\bm{w}_1,\bm{w}_2,\ldots,\bm{w}_{N_r^{\mathrm{RF}}}]$,
respectively.
Because the normalized array responses have constant-modulus entries,
the analog beamformers satisfy the constraints in
\eqref{optimization_tx_constraint} and
\eqref{optimization_rx_constraint}.

\subsection{Lloyd Codebook Construction}

A shared digital precoding codebook
$\mathcal{Q}=\{\bm{c}_1,\ldots,\bm{c}_B\}$ with $B=2^b$
is trained offline and stored at both the transmitter and receiver,
where
$\bm{c}_j\in\mathbb{C}^{N_t^{\mathrm{RF}}\times1}$
and
$\|\bm{c}_j\|_2=1$.
Consequently, each selected codeword can be represented
by a $b$-bit index.

The codebook is obtained offline using the standard Linde--Buzo--Gray vector-quantization procedure~\cite{Linde1980}, which alternates nearest-codeword assignment and normalized centroid updates over unit-norm training vectors until convergence.
If the long-term channel statistics or system configuration change substantially, the shared codebook can be regenerated offline and redistributed to both transceiver ends. No online codebook adaptation is assumed in the present framework.

\subsection{Pilot-Subcarrier Digital Precoder Selection}

Let $k_q$ denote the $q$-th pilot subcarrier, where $q$ indexes the
pilot subcarriers. For each $k_q$, the receiver sequentially selects
$N_s$ codewords from $\mathcal{Q}$ using the corresponding effective
channel.
The residual effective channel is initialized as
$\bm{H}^{(1)}[k_q]=\bm{H}_{\mathrm{eff}}[k_q]$.

At the $t$-th stream-selection step, the receiver selects
\begin{align}
j_{q,t}
=
\arg\max_{j\in\{1,\ldots,B\}}
\left\|
\bm{H}^{(t)}[k_q]\bm{c}_j
\right\|_2^2,
\quad
t=1,\ldots,N_s,
\label{greedy_codeword_selection}
\end{align}
with
$\bm{c}_{q,t}=\bm{c}_{j_{q,t}}$.

Because each codeword is unit norm, the residual effective channel is updated as
$\bm{H}^{(t+1)}[k_q]
=\bm{H}^{(t)}[k_q]
(\bm{I}_{N_t^{\mathrm{RF}}}
-\bm{c}_{q,t}\bm{c}_{q,t}^{H})$.
This sequential projection suppresses the contribution of the
previously selected direction when searching for the subsequent
spatial precoding directions.

After $N_s$ iterations, the pilot-subcarrier digital precoder is
$\bm{V}_q=[\bm{c}_{q,1},\bm{c}_{q,2},\ldots,\bm{c}_{q,N_s}]
\in\mathbb{C}^{N_t^{\mathrm{RF}}\times N_s}$.
The receiver feeds back the $N_s$ codeword indices defining
$\bm{V}_q$.
These pilot-subcarrier precoding directions are subsequently used by the hierarchical interpolation method to construct the digital
precoders for the unreported subcarriers.

\section{Proposed Hierarchical Interpolation}
\label{sec:hierarchical}

Conventional interpolation methods~\cite{He2011,Pande2007}
construct the precoders of unreported subcarriers from a set of
reported pilot precoders according to predetermined interpolation
structures. Although these methods reduce the number of explicitly
reported precoders, they do not explicitly identify the subcarrier
location at which the preferred codeword changes between two
neighboring pilot subcarriers. To address this issue, we propose a
binary-search-based hierarchical interpolation method that adaptively
determines the codeword switching location within each pilot interval.

\subsection{Pilot Arrangement and Single-Transition Model}

Let $M$ denote the pilot spacing. The number of pilot subcarriers is
$N_p=\left\lfloor(K-1)/M\right\rfloor+1$, with locations
$k_q=1+(q-1)M$, $q=1,\ldots,N_p$.
Hence, binary-search interpolation is performed over the
$N_p-1$ intervals between adjacent pilot subcarriers.
If $k_{N_p}<K$, the digital precoder associated with the last
pilot subcarrier is reused for the remaining subcarriers
$k_{N_p}+1,\ldots,K$.

For the $t$-th spatial stream in the $q$-th pilot interval,
let $\bm{v}_q^t$ and $\bm{v}_{q+1}^t$ denote the codewords selected
at pilot subcarriers $k_q$ and $k_{q+1}$, respectively. We define
their relative effective-channel gain as
\begin{align}
\Delta_{q,t}[k]
=
\left\|
\widehat{\bm{H}}_{\mathrm{eff}}[k]
\bm{v}_q^t
\right\|_2
-
\left\|
\widehat{\bm{H}}_{\mathrm{eff}}[k]
\bm{v}_{q+1}^t
\right\|_2,
\label{relative_gain}
\end{align}
where $\widehat{\bm{H}}_{\mathrm{eff}}[k]$ denotes the estimated
effective channel. Under perfect CSI,
$\widehat{\bm{H}}_{\mathrm{eff}}[k]
=
\bm{H}_{\mathrm{eff}}[k]$.

The proposed binary-search procedure assumes that the relative
preference between the two boundary codewords exhibits at most one
transition within a pilot interval. For an interval containing a
switching boundary $\eta_{q,t}$, $\eta_{q,t}$ denotes the last
subcarrier assigned to $\bm{v}_q^t$. The relative gain is then
assumed to approximately satisfy
\begin{align}
\Delta_{q,t}[k]
&\geq 0,
&&
k_q \leq k \leq \eta_{q,t},
\nonumber\\
\Delta_{q,t}[k]
&<0,
&&
\eta_{q,t}<k<k_{q+1}.
\label{single_transition_condition}
\end{align}
This single-transition model is motivated by the frequency correlation among neighboring subcarriers. When multiple preference reversals occur within the same pilot interval, the proposed method provides a single-boundary approximation rather than an exact subcarrier-wise assignment.
Likewise, if an intermediate subcarrier prefers a codeword different from both boundary codewords, the proposed method restricts the assignment to the two boundary candidates and therefore provides an approximation to full subcarrier-wise codeword selection.

\subsection{Binary-Search-Based Hierarchical Interpolation}

If $\bm{v}_q^t=\bm{v}_{q+1}^t$, the same codeword is assigned throughout the interval. Otherwise, a binary search compares the two boundary codewords at each midpoint and recursively retains the preferred half. For power-of-two $M$, the switching boundary is obtained after $\log_2M-1$ comparisons. The complete procedure is summarized in Algorithm~\ref{alg:hierarchical}.

\begin{algorithm}[t]
\caption{\small Proposed Binary-Search-Based Hierarchical Interpolation}
\label{alg:hierarchical}
\SetKwInput{KwData}{Input}
\SetKwInput{KwResult}{Output}

\KwData{
$\{\bm{V}_q\}_{q=1}^{N_p}$,
$\{\widehat{\bm{H}}_{\mathrm{eff}}[k]\}_{k=1}^{K}$,
$\{k_q\}_{q=1}^{N_p}$, $M$, and $N_s$.
}

\KwResult{
$\{\bm{F}[k]\}_{k=1}^{K}$.
}

\For{$q=1$ \KwTo $N_p-1$}{
    \For{$t=1$ \KwTo $N_s$}{
        \eIf{$\bm{v}_q^t=\bm{v}_{q+1}^t$}{
            $\bm{d}_t[k]\gets\bm{v}_q^t$,
            $k=k_q,\ldots,k_{q+1}-1$\;
        }{
            $\zeta_{\mathrm{L}}\gets k_q$,
            $\zeta_{\mathrm{R}}\gets k_{q+1}$\;

            \For{$r=1$ \KwTo $\log_2M-1$}{
                $\zeta_{\mathrm{mid}}
                \gets(\zeta_{\mathrm{L}}+\zeta_{\mathrm{R}})/2$\;

                \eIf{
                $\|
                \widehat{\bm{H}}_{\mathrm{eff}}[\zeta_{\mathrm{mid}}]
                \bm{v}_q^t
                \|_2
                \geq
                \|
                \widehat{\bm{H}}_{\mathrm{eff}}[\zeta_{\mathrm{mid}}]
                \bm{v}_{q+1}^t
                \|_2$
                }{
                    $\zeta_{\mathrm{L}}\gets\zeta_{\mathrm{mid}}$\;
                }{
                    $\zeta_{\mathrm{R}}\gets\zeta_{\mathrm{mid}}$\;
                }
            }

            $\eta_{q,t}
            \gets(\zeta_{\mathrm{L}}+\zeta_{\mathrm{R}})/2$\;

            $\bm{d}_t[k]\gets\bm{v}_q^t$,
            $k=k_q,\ldots,\eta_{q,t}$\;

            $\bm{d}_t[k]\gets\bm{v}_{q+1}^t$,
            $k=\eta_{q,t}+1,\ldots,k_{q+1}-1$\;
        }
    }
}

\If{$k_{N_p}<K$}{
    $\bm{d}_t[k]\gets\bm{v}_{N_p}^t$,
    $t=1,\ldots,N_s$,
    $k=k_{N_p}+1,\ldots,K$\;
}

$\bm{F}[k]\gets
[\bm{d}_1[k],\ldots,\bm{d}_{N_s}[k]]$,
$k=1,\ldots,K$\;

\end{algorithm}

\subsection{Online Complexity and Feedback Requirement}

Let $\mathcal{A}=\{(q,t):\bm{v}_q^t\neq\bm{v}_{q+1}^t\}$ denote the set of stream--interval pairs that require binary search. Each active pair requires $\log_2M-1$ binary-search comparisons. Therefore, the total number of midpoint comparisons is $C_{\mathrm{BS}}=|\mathcal{A}|(\log_2M-1)$. Since each comparison evaluates two candidate codewords, the corresponding number of effective-channel gain evaluations is $2C_{\mathrm{BS}}$.

In the worst case,
$C_{\mathrm{BS}}^{\mathrm{worst}}
=N_s(N_p-1)(\log_2M-1)$.
Hence, the hierarchical interpolation stage requires approximately
$\mathcal{O}(N_s(K/M)\log M)$ gain comparisons, rather than
exhaustively evaluating all intermediate subcarriers.

The feedback consists of two components. First, the receiver reports
the $N_s$ codeword indices at each of the $N_p$ pilot subcarriers,
requiring $B_{\mathrm{pilot}}=N_sN_pb$ bits.
Second, for each active stream--interval pair, the detected
switching location can be represented by the
$\log_2M-1$ binary-search decisions. Therefore,
$B_{\mathrm{sw}}=|\mathcal{A}|(\log_2M-1)$ bits.

The total and worst-case feedback requirements are, respectively,
\begin{align}
B_{\mathrm{prop}}
&=
N_sN_pb+|\mathcal{A}|(\log_2M-1),
\label{proposed_feedback}\\
B_{\mathrm{prop}}^{\mathrm{worst}}
&=
N_sN_pb+N_s(N_p-1)(\log_2M-1).
\label{worst_feedback}
\end{align}

\subsection{Linear Digital Combining}

For minimum mean-square error (MMSE) detection, let
$\bm{A}[k]=\widehat{\bm{H}}_{\mathrm{eff}}[k]\bm{F}[k]$ and use
$\bm{W}_{\mathrm{MMSE}}^{H}[k]
=\bm{A}^{H}[k]
\left(
\bm{A}[k]\bm{A}^{H}[k]
+\frac{N_s\sigma_n^2}{P}
\bm{W}_{\mathrm{RF}}^{H}\bm{W}_{\mathrm{RF}}
\right)^{\dagger}$,
where $(\cdot)^{\dagger}$ denotes the Moore--Penrose pseudoinverse.

\section{Simulation Results}
\label{sec:simulation}

We consider a single-user mmWave MIMO-OFDM system with $N_s=2$, $N_t=128$, $N_r=32$, $N_t^{\mathrm{RF}}=16$, $N_r^{\mathrm{RF}}=12$, $K=2048$, $M=128$, $L=24$, and $D=32$. Among the $K=2048$ OFDM subcarriers, $N_p=16$ subcarriers are used as pilot subcarriers and the remaining $2032$ subcarriers carry data. The shared digital precoding codebook contains $B=32$ unit-norm codewords, corresponding to $b=5$ feedback bits per codeword index. The carrier frequency is $60$ GHz, and the geometric wideband mmWave channel model follows \cite{Ayach2014,Alkhateeb2016}. All reported results are averaged over 1000 independent channel realizations with independently generated transmitted signals and noise.

For $K=2048$ and $M=128$, $N_p=16$, yielding 15 pilot intervals, and each active pair requires $\log_2 128-1=6$ comparisons.

We compare the proposed method with Gaussian interpolation~\cite{He2011}, geodesic interpolation~\cite{Pande2007}, simple clustering, and SNR-maximizing clustering~\cite{Wu2009}. All schemes use the same system configuration, pilot arrangement, and codebook size.

\subsection{Feedback Overhead Analysis}

We quantify only digital-precoder feedback under an error-free, delay-free feedback link; channel estimation, beam training, synchronization, feedback errors, and latency are excluded. Accordingly, the reported reduction refers to digital-precoder feedback rather than total signaling overhead.
Full-grid per-subcarrier feedback requires
$B_{\mathrm{conv}}=KN_sb=20480$ bits.

For the proposed scheme, the receiver first reports the $N_s$ codeword indices on each of the $N_p$ pilot subcarriers. This requires $B_{\mathrm{pilot}}=N_sN_pb=160$ bits.

In addition, for each stream--interval pair whose two boundary codewords differ, the receiver reports the binary branch decisions generated during the hierarchical search. Since each active pair requires $\log_2M-1$ binary-search levels, the corresponding feedback requires $\log_2M-1$ bits. The total feedback is given by~\eqref{proposed_feedback}.

In the worst case, all adjacent pilot codewords differ for all
spatial streams.
Using~\eqref{worst_feedback}, the worst-case feedback becomes
$B_{\mathrm{prop}}^{\mathrm{worst}}
=2(16)(5)+2(15)(6)=340$ bits.
Consequently, even under this worst-case switching condition, the
feedback reduction relative to conventional per-subcarrier feedback is $1-340/20480=98.34\%$.
This value is referenced specifically to full-grid per-subcarrier feedback and should not be interpreted as the feedback reduction relative to the reduced-feedback interpolation or clustering baselines.

\subsection{Spectral Efficiency Performance}

\begin{figure}[t]
\centering
\includegraphics[width=0.42\textwidth]{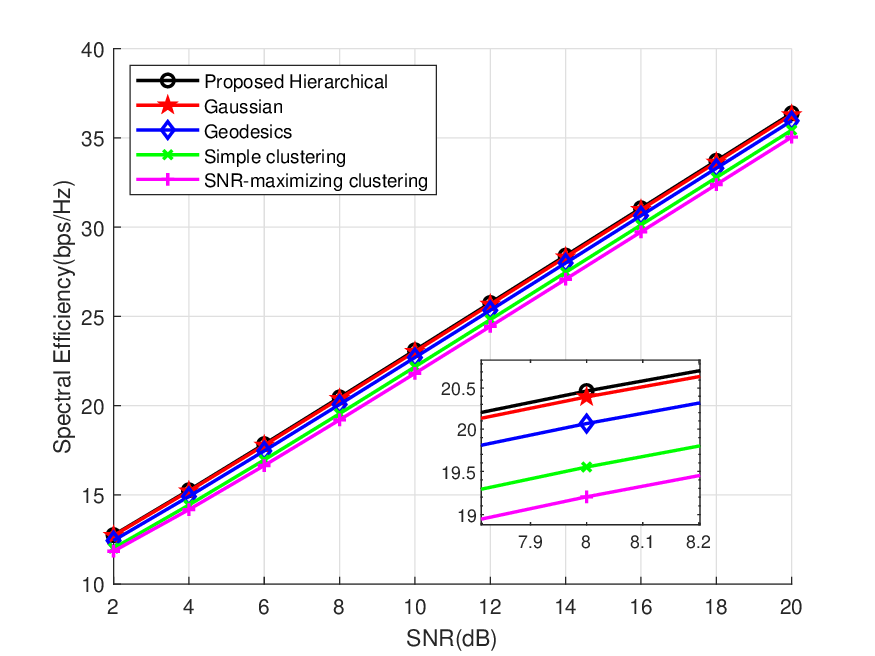}
\caption{\small Spectral efficiency comparison of different precoding schemes.}
\label{Spectral}
\end{figure}

Fig.~\ref{Spectral} shows that the proposed method achieves spectral efficiency comparable to the competing schemes under the considered configuration, illustrating a favorable performance--complexity trade-off without implying uniform superiority over all baselines.

\subsection{BER Performance}

\begin{figure}[t]
\centering
\includegraphics[width=0.42\textwidth]{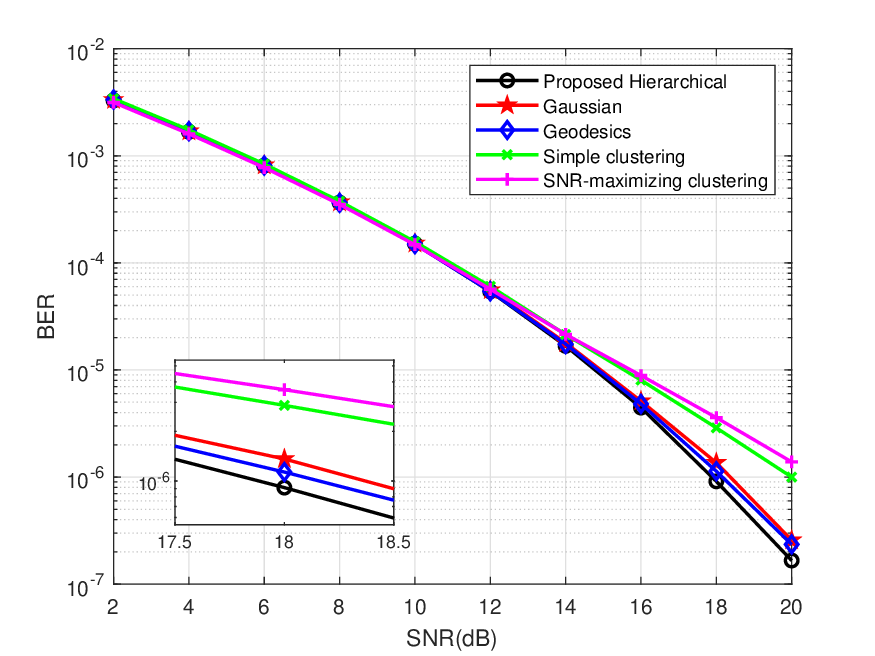}
\caption{\small Average BER performance of different precoding schemes.}
\label{Sim_BER}
\end{figure}

Fig.~\ref{Sim_BER} compares the BER performance of the considered
precoding schemes under quadrature phase-shift keying (QPSK) modulation and perfect CSI.

The proposed method achieves competitive BER across the considered SNR range while avoiding exhaustive subcarrier-wise codeword comparisons. These results indicate that the hierarchical search preserves reliable error performance with low online search complexity under the considered configuration.

\subsection{Computational Complexity}

We compare the average runtime over 1000 independent channel realizations.
The reported runtime measures the online precoding/interpolation stage
and excludes the offline Lloyd codebook training.
\begin{table}[t]
\centering
\caption{\small Computation Time of Different Precoding Algorithms}
\setlength{\tabcolsep}{3pt}
\renewcommand{\arraystretch}{0.9}
\begin{tabular}{|l|c|}
\hline
Method & Average time (s) \\
\hline
SNR-maximizing clustering~\cite{Wu2009}
& $2.6 \times 10^{-1}$ \\
\hline
Simple clustering~\cite{Wu2009}
& $2.3 \times 10^{-3}$ \\
\hline
Gaussian interpolation~\cite{He2011}
& $5.2 \times 10^{-3}$ \\
\hline
Geodesic interpolation~\cite{Pande2007}
& $1.5 \times 10^{-3}$ \\
\hline
Proposed hierarchical interpolation
& $\bm{6.7 \times 10^{-4}}$ \\
\hline
\end{tabular}
\label{Sim_Comp_Time}
\end{table}

As shown in Table~\ref{Sim_Comp_Time}, the proposed method achieves
the lowest runtime. Gaussian interpolation~\cite{He2011} and geodesic
interpolation~\cite{Pande2007} require approximately $7.8$ and $2.2$
times longer runtime, respectively, while SNR-maximizing
clustering~\cite{Wu2009} requires over two orders of magnitude more
time. This agrees with the
$\mathcal{O}(N_s(K/M)\log M)$ complexity derived in
Section~\ref{sec:hierarchical}.

\subsection{Performance Under Imperfect CSI}

We finally evaluate robustness under imperfect effective CSI, modeled as
$\widehat{\bm{H}}_{\mathrm{eff}}[k]
=\sqrt{\rho}\bm{H}_{\mathrm{eff}}[k]
+\sqrt{1-\rho}\bm{H}_{\mathrm{err}}[k]$,
where $\rho\in[0,1]$ denotes the CSI quality.
The entries of $\bm{H}_{\mathrm{err}}[k]$ are independently generated
as zero-mean complex Gaussian variables and normalized to have the same
average power as $\bm{H}_{\mathrm{eff}}[k]$.
Here, $\rho=1$ represents perfect CSI.

\begin{figure}[t]
\centering
\includegraphics[width=0.45\textwidth]{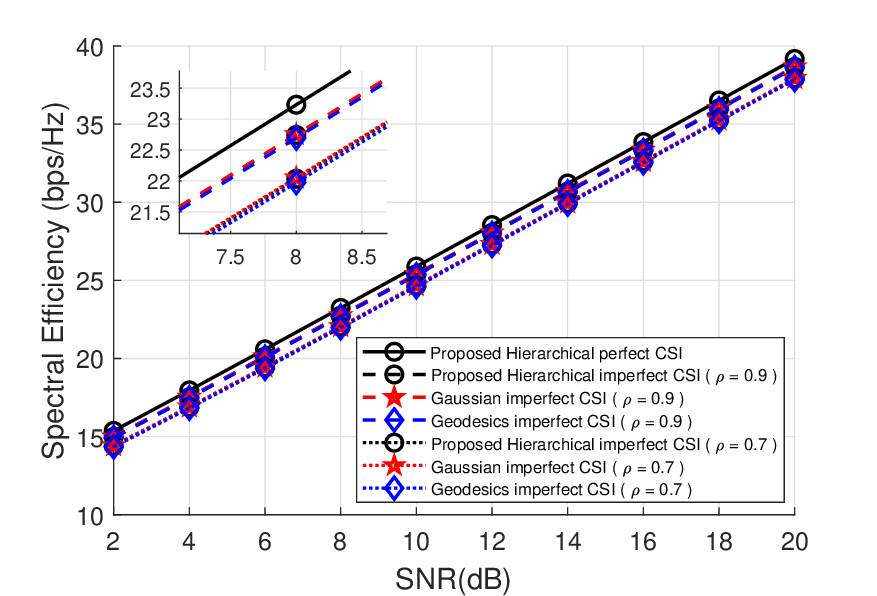}
\caption{\small Spectral efficiency comparison under perfect and imperfect CSI conditions.}
\label{ImperfectCSI}
\vspace{-0.1in}
\end{figure}

Fig.~\ref{ImperfectCSI} compares the spectral efficiency of the
proposed, Gaussian-interpolation, and geodesic-interpolation schemes
under different CSI-quality levels.

As CSI quality degrades, spectral efficiency decreases because effective-channel perturbations affect codeword selection and digital combining. The proposed method remains competitive for moderate errors, since quantization can make the selected codeword and switching boundary remain unchanged unless the perturbation alters the relative preference between candidate codewords.

The present evaluation is limited to one configuration with fixed pilot spacing, codebook size, RF-chain numbers, and propagation parameters. The reported results should therefore be interpreted within this operating regime; sensitivity to these parameters and multiple-transition conditions is left for future work.

\vspace{-0.1in}
\section{Conclusion}

We proposed a reduced-feedback hybrid precoding framework combining dominant-path analog beamforming, Lloyd-codebook-based digital precoding, and hierarchical interpolation. Under the considered configuration, it reduces worst-case feedback by $98.34\%$ relative to full-grid per-subcarrier feedback while maintaining competitive spectral efficiency and BER. Extensions to multi-transition and broader parameter settings are left for future work.

\end{document}